# STEM: Soft Tactile Electromagnetic Actuator for Virtual Environment Interactions

Heeju Mun[1], Seunggyeom Jung[1], Seung Mo Jeong[1],
David Santiago Diaz Cortes[1], and Ki-Uk Kyung[1]

[1] Mechanical Engineering, Korea Advanced Institute of Science and Technology (KAIST), Daejeon, South Korea

(Email: kyungku@kaist.ac.kr)

**Abstract ---** The research aims to expand tactile feedback beyond vibrations to various modes of stimuli, such as indentation, vibration, among others. By incorporating soft material into the design of a novel tactile actuator, we can achieve multi-modality and enhance the device's wearability, which encompasses compliance, safety, and portability. The proposed tactile device can elevate the presence and immersion in VR by enabling diverse haptic feedback such as, force indentation, vibration and other arbitrary force outputs. This approach enables the rendering of haptic interactions with virtual objects, such as grasping of aa 3D virtual object to feel its stiffness – action that was difficult to achieve using widely adopted vibrotactile motors.

**Keywords:** electromagnetic soft actuator, VR haptic interaction, wearable device

## 1 INTRODUCTION

With continuous advancement of virtual reality (VR), research into wearable haptic devices that simulate tactile interactions for an immersive experience in virtual environments is ongoing.[1-2] Humans interact with the physical world through haptics, highlighting the importance of wearable haptic feedback devices in VR technologies for enhancing immersion, and interaction. By providing only tactile feedback, users can directly manipulate three-dimensional virtual objects, enhancing overall performance by reducing cognitive effort and dependence on vision when completing certain tasks.[3-4]

Wearability of the device is one of the critical factors that influences the immersiveness in VR environment. Rigid and bulky devices with low wearability, not only pose safety threats to the wearer but also disrupt the cognitive connection between virtual and physical interactions due to high impedance.[5] In this aspect, small and compact vibrotactile haptic actuators, such as LRAs or coin motors are often adopted.[6-7] They are capable of providing strong vibrations at specific frequency, which is enough for notifying the users about certain interactions. To generate feedback in a much less dynamic state, small-scale motors are utilized. Spatial haptic surfaces including the virtual surface bumps and the surface patterns are rendered using linear motors that produce indentational force at the finger tips to deform the skin beneath.[8] Another means of delivering the pressing sensation is using a servo motor, encased in a thimble structure.[9-10] While these approaches can deliver single mode of tactile feedback, studies to design a device with increased degree-of-freedom are gaining interests. Common approach is to use multiple actuators or motors that generate different modes of tactile stimuli, such as pressure, vibration, and shear.[11-12] However, this attribute comes at the cost of enlarged system, thereby deteriorating the user's accessibility in using the device.

Recently, soft materials are actively leveraged in designing wearable haptic actuators as they offer potential solutions to achieve compliant and lightweight devices. Additionally, soft tactile actuators enable conformal contact with the finger curvature, which is crucial for effective feedback transmission. Various approaches have been proposed for developing such actuators. Dielectric and electrostatic force-based actuators exhibit fast responses with high energy density at high voltage input (~kV).[13-14] Pneumatic actuation, on the other hand, demonstrates pronounced force output (~N) with limited bandwidth in the range of tens of Hertz, and the bulky air compressor restricts the movement.[15-16] Despite an innovative approach of actuation base on soft functional material, it must overcome significant challenges, including relatively weak force output, robustness, and feasibility.[5]

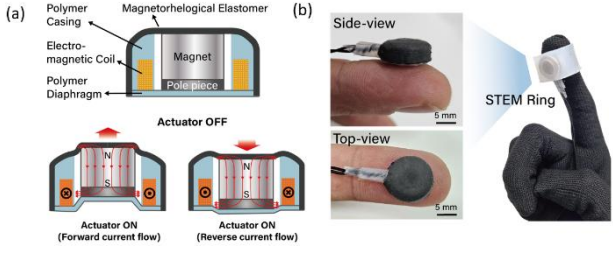

Fig.1 (a)Schematic of the STEM actuator and the working mechanism. (b) Fabricated actuator and the haptic glove with the STEM Ring

Alternatively, instead of using soft material as the primary actuating element, they can be incorporated as a passive structure, serving as the encasing or an energy-storing element. Yu *et al*. and Li *et al*. have proposed a miniaturized electromagnetic tactile actuator utilizing flexible film as the elastic structure.[17-18] As their focus was on miniaturizing the actuator size with minimum power input, the device was designed for vibration at a specific resonance frequency exclusively.

Herein, the study proposes a soft tactile electromagnetic (STEM) actuator that can be finger-worn and deliver tactile feedback not just of mere vibrations but also of force, impulse, and arbitrary forms of feedback. By incorporating soft material as an energy-storing element and an encasing structure of the actuator, we achieve out-of-plane deformation of the tactile actuator in response to any input signal waveform with enhanced wearability of the haptic device. With the developed hardware, we aim to render the haptic interaction in virtual environments, such as pressing a button, and grasping of an object with realism.

## 2 Soft Tactile Electromagnetic Actuator Design

### 2.1 Actuator Structure Design

The STEM actuator operates similarly to a voice coil actuator. It consists of an electromagnetic coil embedded in a silicone polymer, and a permanent NdFeB magnet suspended inside the electromagnetic coil with the support of polymer diaphragm (see Fig. 1a). The ferrimagnetic pole-piece attached to the magnet amplifies and directs the magnetic flux lines perpendicularly to the coil's current. The magnetorheological elastomer(MRE), a composite of polymer, and magnetic particles, minimizes magnetic flux leakage, enhancing actuation efficiency. Upon actuation, current through the coil generates a Lorentz Force, pushing the magnet upward.

The design parameters of the coil and the magnet, such as magnet height, $h_{mag}$ and the width of the coil, $w_{coil}$ are determined using the FEM simulation results. The objective function is set as Equation (1) where $F, P$ and $m$ represent the magnetic force output, input power and mass of the magnet respectively.

$$f(h_{mag}, w_{coil}) = \frac{F}{\sqrt{Pm}} \qquad (1)$$

Based on the FEM result, we set the magnet as 4 mm and the coil width as 2 mm, considering the overall volume of the actuator structure. The coil thickness is set to 3 mm, and the magnet radius is set to 2 mm. The fabricated STEM actuator has a finger-scale size, with a diameter of 11 mm and a thickness of 6 mm (See Fig. 1b).

Another crucial design factor in determining the actuator's performance is the volume percentage of Carbonyl Iron Particles (CIP) in the polymer matrix. A higher concentration of CIP allows the MRE to contain the magnetic flux, minimizing leakage. To experimentally validate this effect, different designs with varying volume concentration (Table. 1) were prepared and tested.

Table 1 Concentration of CIP in four different designs

| Design | CIP vol. concentration |
| --- | --- |
| Design 1 | 0 % |
| Design 2 | 10 % |
| Design 3 | 20 % |
| Design 4 | 30 % |

### 2.2 Actuator Performance

The force output of the actuator is measure using loadcell attached to the vertical motor stage. Initial load of 50 mN is applied to set a similar condition to when the actuator is in contact with the skin. Step input voltage with increasing magnitude of 0.2 V is given, and the static force output is measured for 30 seconds. Four measurements are taken for four samples of each design. The averaged static force output for different designs with increasing input voltage are plotted in Fig. 2a. The solid line indicates the averaged value whereas the colored region indicates the standard deviation. From the result, we can identify that with increasing concentration of CIP particles in the polymer diaphragm, output force can be enhanced. This observation thus supports the idea of minimizing magnetic flux leakage using magnetic material as a part of the structure.

The force response to sinusoidal voltage input with varying frequency and amplitude of 3 V, 5 V and 7 V was measured. The maximum current flow at each voltage

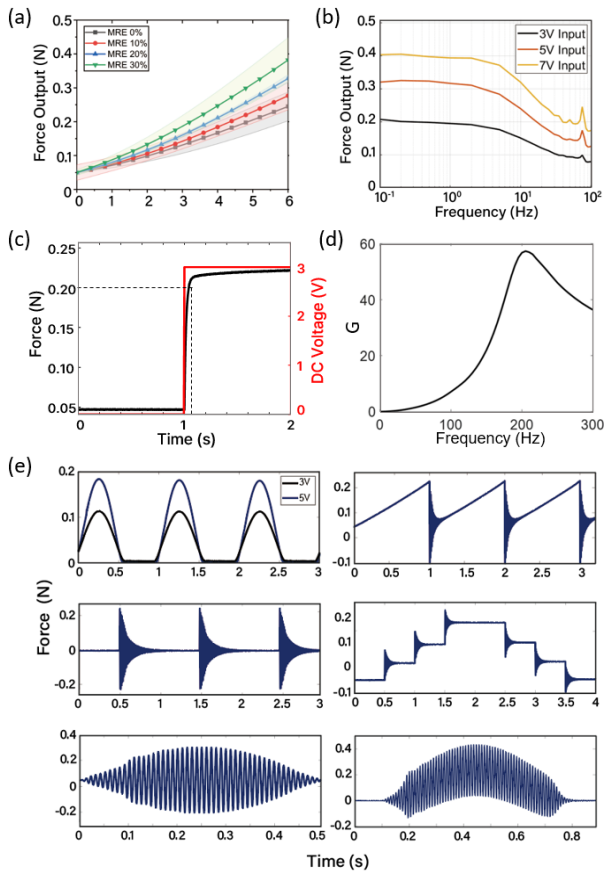

a) Blocked force output with increasing input voltage for different CIP volume concentration, b) Force response for varying frequency input signals with increasing voltage amplitude, c) Step force response of the actuator to observe the response time, d) Output acceleration response with varying input frequency signals, e) Output response for arbitrary input signals including sinusoidal, ramp, impulse, and so on.

input was 350 mA, 600 mA, and 750 mA, respectively. The result is shown in Fig. 2b, where the output is significant in the low-frequency region around 10 Hz. The actuator is capable of producing a force of up to 0.4 N with an input signal of 7 V. To investigate the response time, the force was measured under a step input signal. The time taken for the actuator to reach 90% of the maximum force output was 44.6 ms (see Fig. 2c), demonstrating a fast response suitable for reliable haptic rendering. Gravitational acceleration (G, 9.8 m/s²) was also measured with varying frequencies of sinusoidal input voltage with an amplitude of 3 V. The result is shown in Fig. 2d, where the output acceleration exceeds 1 G starting from 40 Hz. The resonance was found at 210 Hz, where the maximum acceleration of 58 G was observed.

The actuator was also able to produce arbitrary force output, as shown in Fig. 2e. This capability arises from the fact that the force is generated by the out-of-plane deformation of the polymer diaphragm in response to the input signal. As a result, the actuator exhibits multi-modal tactile feedback properties. However, a drawback is the heat generated due to the high current flow through the coil. A thermal camera was used to measure the temperature rise, which reached as high as 40°C under prolonged sinusoidal input with a frequency of 100 Hz and an amplitude of 3 V for 100 seconds. Given that haptic interactions typically last no longer than 5 seconds, it is unlikely to cause harm to the skin.

## 3 Wearable Haptic Device System

The developed STEM actuator is mounted on the user's finger using a soft structure known as the Haptic Ring (see Fig. 1b). This ring adapts to different finger sizes and is attached to a stretchable glove. The control system consists of a microcontroller unit (Teensy 4.0), a PC, a head-mounted display (MetaQuest 3), and a motor driver (Adafruit HR8833). The PC collects the 3-dimensional coordinates of the user's finger from the MetaQuest and sends the appropriate voltage signal to the motor driver, which generates the high input current needed for actuation.

Depending on the virtual objects the user interacts with, different tactile feedback modes, such as those shown in Fig. 2e, can be delivered. Thanks to the actuator's ability to provide multimodal tactile feedback, we can potentially render haptic interactions with deformable virtual objects, allowing the user to feel variations in stiffness or softness. This type of haptic rendering requires a glove capable of producing significant force and stroke, which is achievable with the proposed actuator design.

## 4 Conclusion

In this study, a novel electromagnetic soft actuator designed to deliver multi-modal tactile feedback to the user is proposed. The actuator's compact size enables it to be conformally worn on the fingertip, thereby reducing the overall weight and size of the haptic glove system. The actuator is capable of producing not only vibrations but also force indentations and other arbitrary force outputs. This versatility allows for rich haptic interaction in VR environments, providing more immersive and realistic experiences for the user. By simulating varying levels of stiffness, softness, and dynamic tactile feedback, the STEM actuator enables a variety of haptic interactions that were difficult to achieve with current commercial haptic gloves. In the future, this work could be extended to develop an array structure to transmit a large amount of information with a multi-modal haptic system.

ACKNOWLEDGEMENT

This work is supported by Culture, Sports, and Tourism R&D Program through the Korea Creative Content Agency grant funded by the Ministry of Culture, Sports, and Tourism in 2024 (Project Name: Development of Barrier Free Technology for the Creation and Collaboration of Choreography for Dancers with Auditory/Visual Impairment, Project Number: RS-2021-KC000738, Contribution Rate: 100%)